\documentclass[a4paper]{spie}  
\usepackage[]{graphicx}

\newcommand{\simPOPS}{\texttt{sim2GFT}}
\newcommand{\superscript}[1]{\ensuremath{^{\textrm{#1}}}}

\def\captionof#1#2{{\def\@captype{#1}#2}}

\title{The Planar Optics Phase Sensor: a study for the VLTI 2\superscript{nd} Generation Fringe Tracker} 

\author{Nicolas Blind\supit{a}, Jean-Baptiste LeBouquin\supit{a}, Olivier Absil\supit{b}, Mazen Alamir\supit{e}, Jean-Philippe Berger\supit{c}, Denis Defr\`ere\supit{b}, Philippe Feautrier\supit{a}, François H\'enault\supit{d}, Laurent Jocou\supit{a}, Pierre Kern\supit{a}, Thomas Laurent\supit{b}, Fabien Malbet\supit{a}, Denis Mourard\supit{d}, Karine Rousselet-Perrault\supit{a}, Alain Sarlette\supit{f}, Jean Surdej\supit{b}, Nassima Tarmoul\supit{d}, Eric Tatulli\supit{a}, Lionel Vincent\supit{a, e}
\skiplinehalf
\supit{a} Laboratoire d'Astrophysique de Grenoble (LAOG), UMR 5571, BP 53, 38041 Grenoble cedex 9, France; \\
\supit{b} Institut d'Astrophysique et de G\'eophysique de Li\`ege (IAGL), University of Li\` ege, B-4000 Sart Tilman, Belgium; \\
\supit{c} European Southern Observatory, Santiago, Chile; \\
\supit{d} Laboratoire H. FIZEAU, Observatoire de la C\^ote d'Azur (OCA), UMR 6525, Avenue Copernic, 06130 Grasse, France;\\
\supit{e} GIPSA-Lab, UMR 5216, BP 46, 38402 Grenoble cedex, France.\\
\supit{f} Dept of Electrical Engineering and Computer Science, University of Li\` ege, B-4000 Sart Tilman, Belgium.
}

\authorinfo{Further author information: (Send correspondence to N.B.)\\
                   N.B.: E-mail: nicolas.blind@obs.ujf-grenoble.fr, Telephone: + 33 4 76 63 57 30\\}

\begin{document}
  \maketitle

\begin{abstract}
In a few years, the second generation instruments of the Very Large Telescope Interferometer (VLTI) will routinely provide observations with 4 to 6 telescopes simultaneously. To reach their ultimate performance, they will need a fringe sensor capable to measure in real time the randomly varying optical paths differences. A collaboration between LAOG (PI institute), IAGL, OCA and GIPSA-Lab has proposed the Planar Optics Phase Sensor concept to ESO for the 2$^{nd}$ Generation Fringe Tracker. This concept is based on the integrated optics technologies, enabling the conception of extremely compact interferometric instruments naturally providing single-mode spatial filtering. It allows operations with 4 and 6 telescopes by measuring the fringes position thanks to a spectrally dispersed  ABCD method. We present here the main analysis which led to the current concept as well as the expected on-sky performance and the proposed design.

\end{abstract}



\section{INTRODUCTION}

With the arrival of the GRAVITY and MATISSE instruments after 2012, and VSI after 2015, the VLTI will routinely work with 4 or possibly 6 telescopes simultaneously. As ground-based interferometric devices, they will be extremely sensitive to the atmospheric turbulence and in particular to the randomly varying optical path difference (OPD) between the telescopes. The random movement of the fringes it involves quickly blurs the interferometric signal, preventing for integration time longer than the coherence time of the atmosphere, typically a few tens of milliseconds in the infrared. The limiting magnitude and precision of ground-based interferometers are therefore dramatically reduced. To reach their ultimate performance, it is then mandatory to dispose of a fringe tracker, that is a device measuring in real time the position of the fringes within a fraction of wavelength and finally stabilizing them.

In the context of the next generation of interferometric instruments, the 2$^{nd}$ Generation Fringe Tracker (2GFT) is intended to cophase an array up to 6 telescopes, formed of Auxiliary and/or Unit Telescopes (ATs and UTs respectively). The Planar Optics Phase Sensor (POPS) is a fringe sensor  proposition for the 2GFT resulting from the collaboration between LAOG (PI institute), IAGL, OCA and GIPSA-Lab. 
We present here the results of the related system studies. They aim at defining the most efficient conceptual design within the VLTI environment, providing precise measurements of the fringe position as well as robust operations, that is a stable fringe tracking whatever the observationnal or atmospheric conditions. To do so, we propose 4-telescope and 6-telescope (4T and 6T hereafter) concepts based on the integrated optics (IO) technology, providing a single-mode spatial filtering as well as pairwise co-axial combinations of the beams, and coding the fringes with a spatial ABCD estimator. These system choices are based on theoretical studies and realistic simulations as well as experience gained in operation with real interferometers, especially VINCI at VLTI\cite{lebouquin_2004, lebouquin_2006} , IONIC-3TH at IOTA\cite{berger_2003}~, and the fringe-sensor FINITO\cite{lebouquin_2008} at VLTI.

We briefly summarize the results of a theoretical analysis of modal filtering influence on the phase measurement in section \ref{part:IO_choice}. The best conceptual architectures for 4- and 6-telescope fringe sensors are determined in relation to their sensitivity, robustness and some additional operational considerations in section \ref{part:combination_choice}. The performance of the most common phase and group delay estimators are then compared in part \ref{part:sensing_concepts}, taking into account atmospheric disturbances. The instrument conceptual design and the results of realistic simulations for the 4T concept are finally presented in sections \ref{part:proposed_concept} and \ref{part:expected_perf} respectively.

	\section{CHOICE OF THE INTEGRATED OPTICS} \label{part:IO_choice}

 The initial idea of POPS is to propose an instrument based on the IO technology. IO components provide very compact interferometric instruments, only a few centimeters long, with low sensitivity to external constraints (vibrations or temperature fluctuations) and they simplify the alignement procedure. Transmission of IO is however lower than their equivalent in bulk optics, by $20$ to $30\%$ for a co-axial pairwise combination for instance, which corresponds to a limiting magnitude $0.3$\,mag lower.

additionally, this technology provides by design a single-mode (SM hereafter) propagation of light, which spatially filters the incoming wavefront and selects its coherent part. Thanks to a theoretical description of the signal, we compare the performance of multi-mode (MM) and SM interferometry for the estimation of the phase (see Tatulli et al\cite{tatulli_2010} (these proceedings) for details). The conclusion of this study is that despite the loss of flux occurring when injecting the light into the SM component (i.e. SM fibers, integrated optics), spatial filtering globally improves fringe sensing performance. 
This study also demonstrates SM interferometry is more efficient for off-axis fringe tracking and equally efficient for the stability of the beams injection. 

Consequently the POPS concept, based on single-mode integrated optics, is a great alternative to classical bulk concepts, from the performance and operational points of view.

\section{THE COMBINATION CONCEPT} \label{part:combination_choice}

A fringe tracker can cophase a N-telescope interferometer by measuring only N-1 OPDs, thanks to N-1 baselines. However because of the noisy measurements and varying observationnal conditions during a night, it can be profitable to dispose of some information redundancy with additional baselines. It is then possible to retrieve the phase on a baseline with several independant ways, insuring a better fringe tracking stability.

All-in-one multi-axial schemes are therefore interesting concepts since they provide all the possible baselines. However they are not currently compatible with the specific requirements of fringe tracking, that is very fast measurements and high  limiting magnitudes. Because of the high number of pixels they require (95 in the 4T case) and of the performance of available detectors (Hawaii II RG, with a read-out noise of $10e^-$/pixel and a low acquisition frequency), their performance is limited in this context. A detailed discussion of the properties and ultimate performance of IO multi-axial combination shemes is presented in Ref.~\citenum{tarmoul_2010} (these proceedings).

		\subsection{Study of the co-axial pairwise schemes}

\begin{figure}[t!]
	\centering
		\begin{tabular}{cccccc}
			\includegraphics[width=0.14\textwidth]{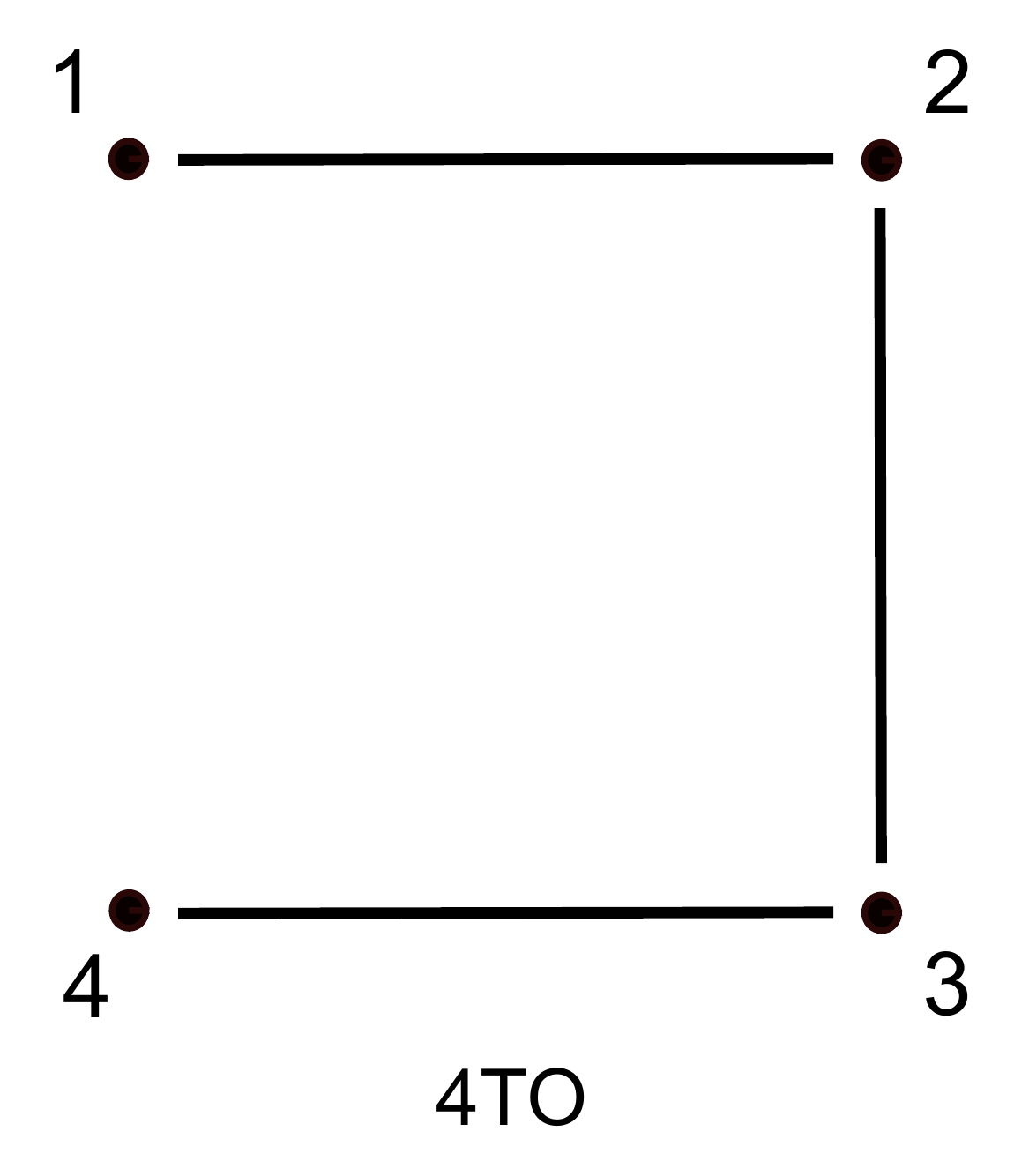}     &
			\includegraphics[width=0.14\textwidth]{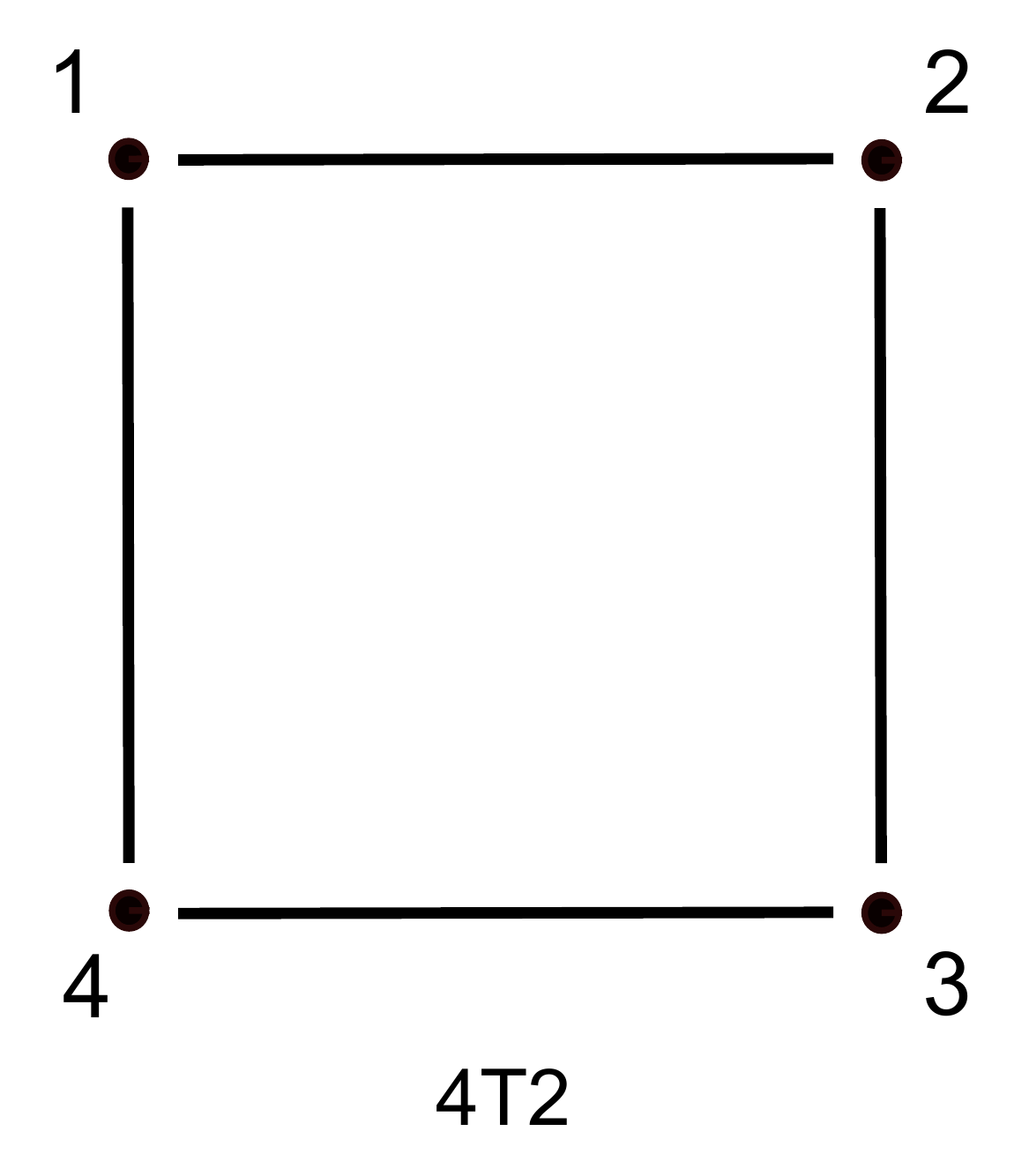}   &
			\includegraphics[width=0.14\textwidth]{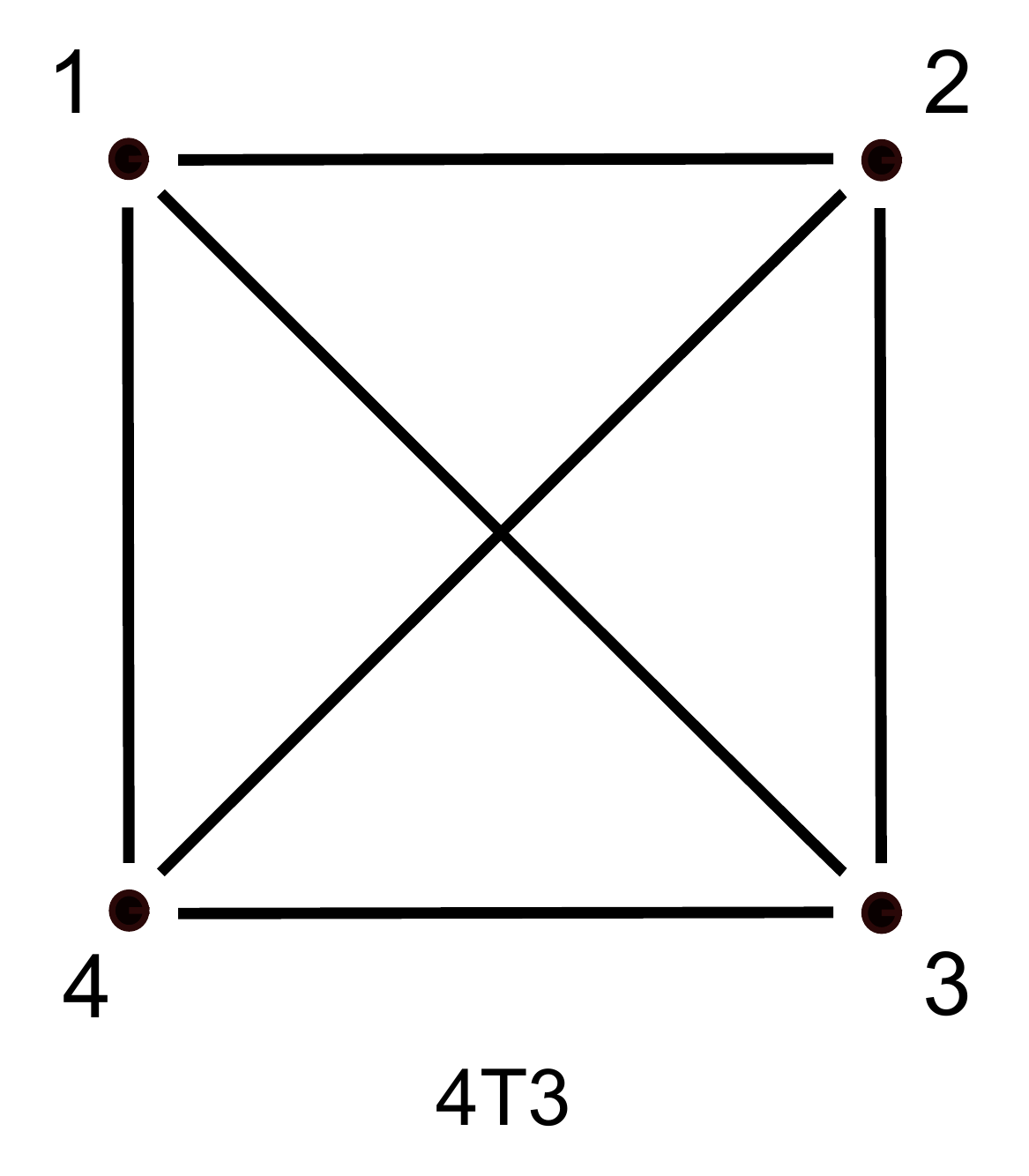}   &
		                                      					                                               &
		                                                        					                           &
		                                                                                      \\
			\includegraphics[width=0.14\textwidth]{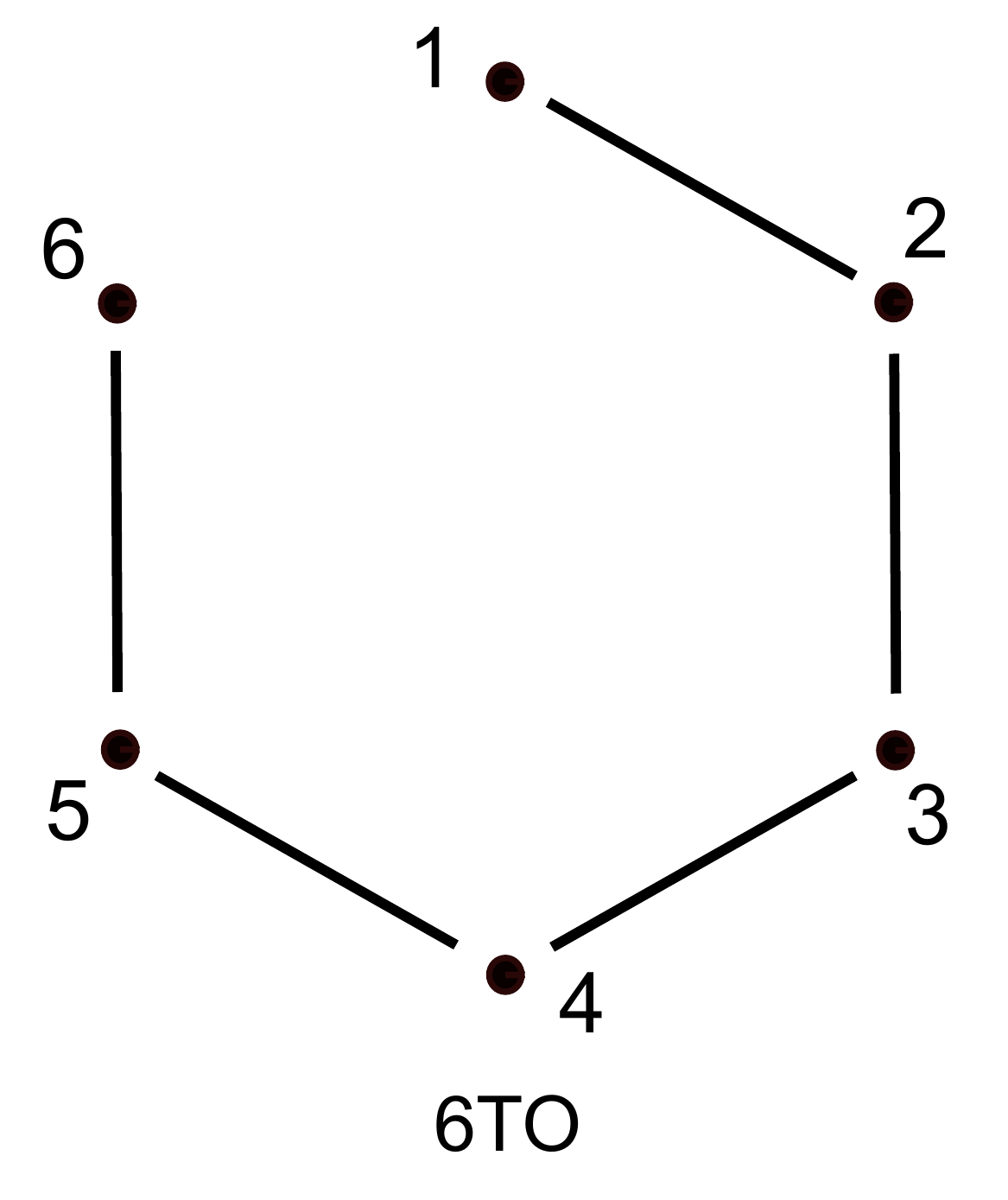}     &
			\includegraphics[width=0.14\textwidth]{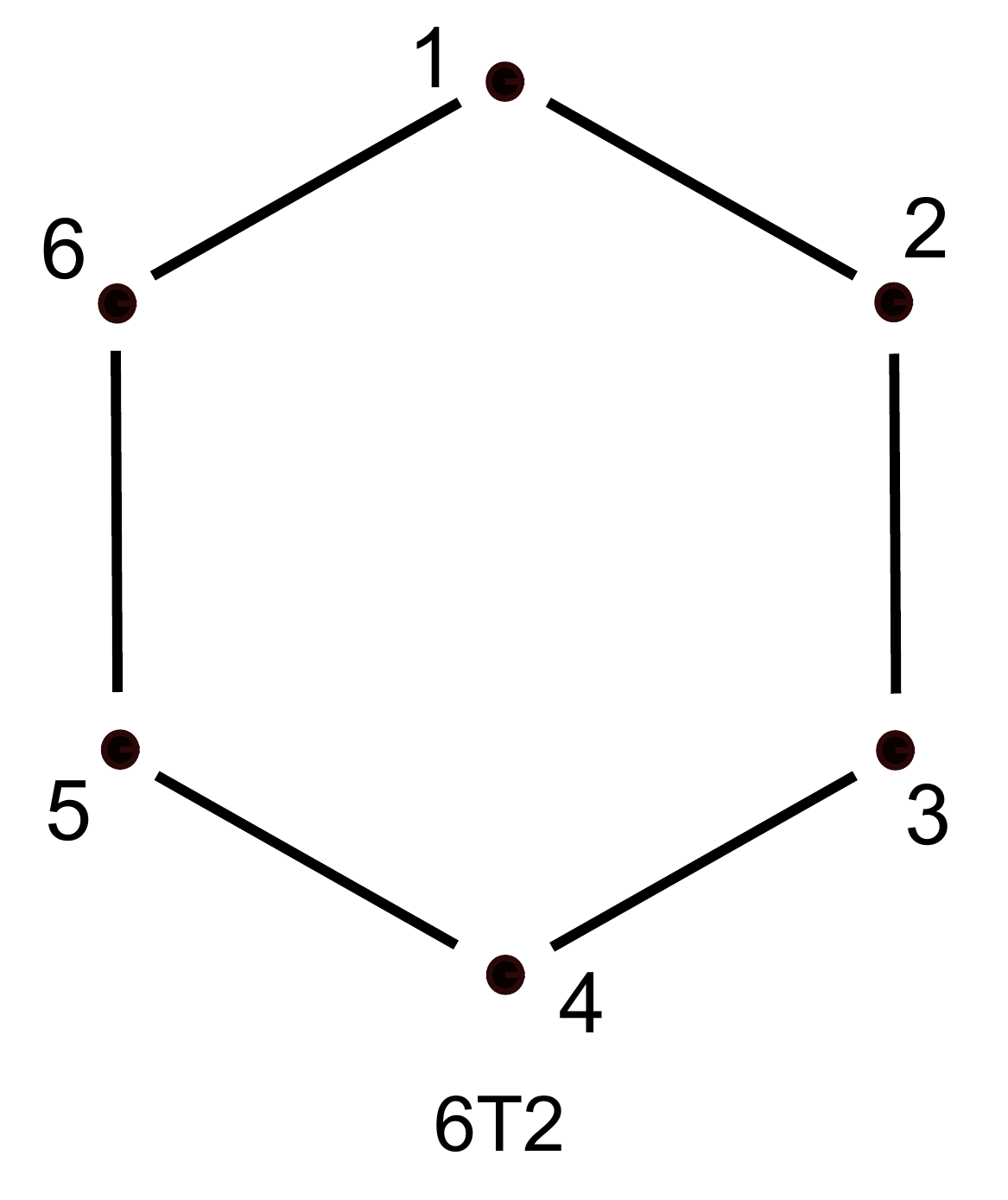}     &
			\includegraphics[width=0.14\textwidth]{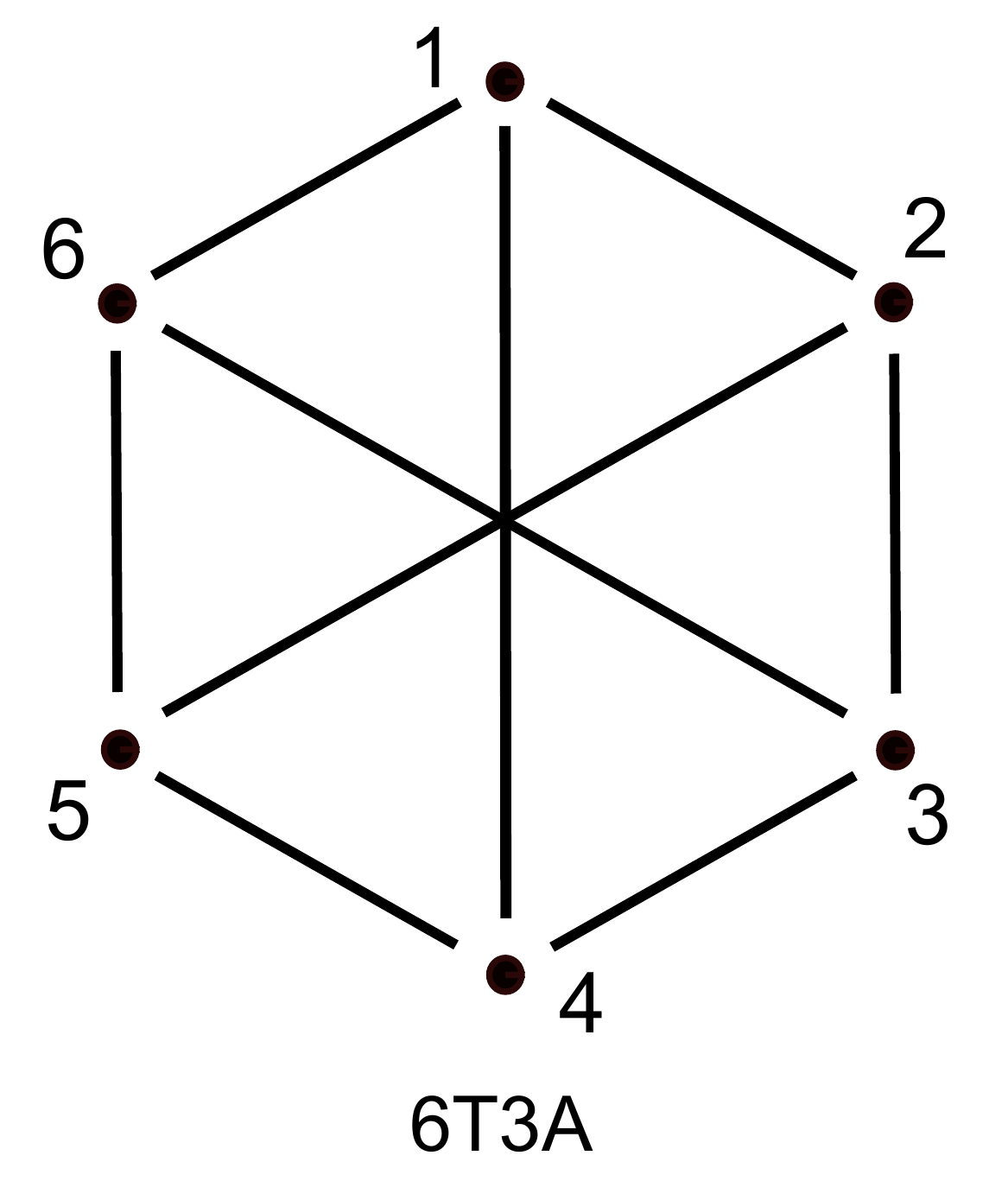}     &
			\includegraphics[width=0.14\textwidth]{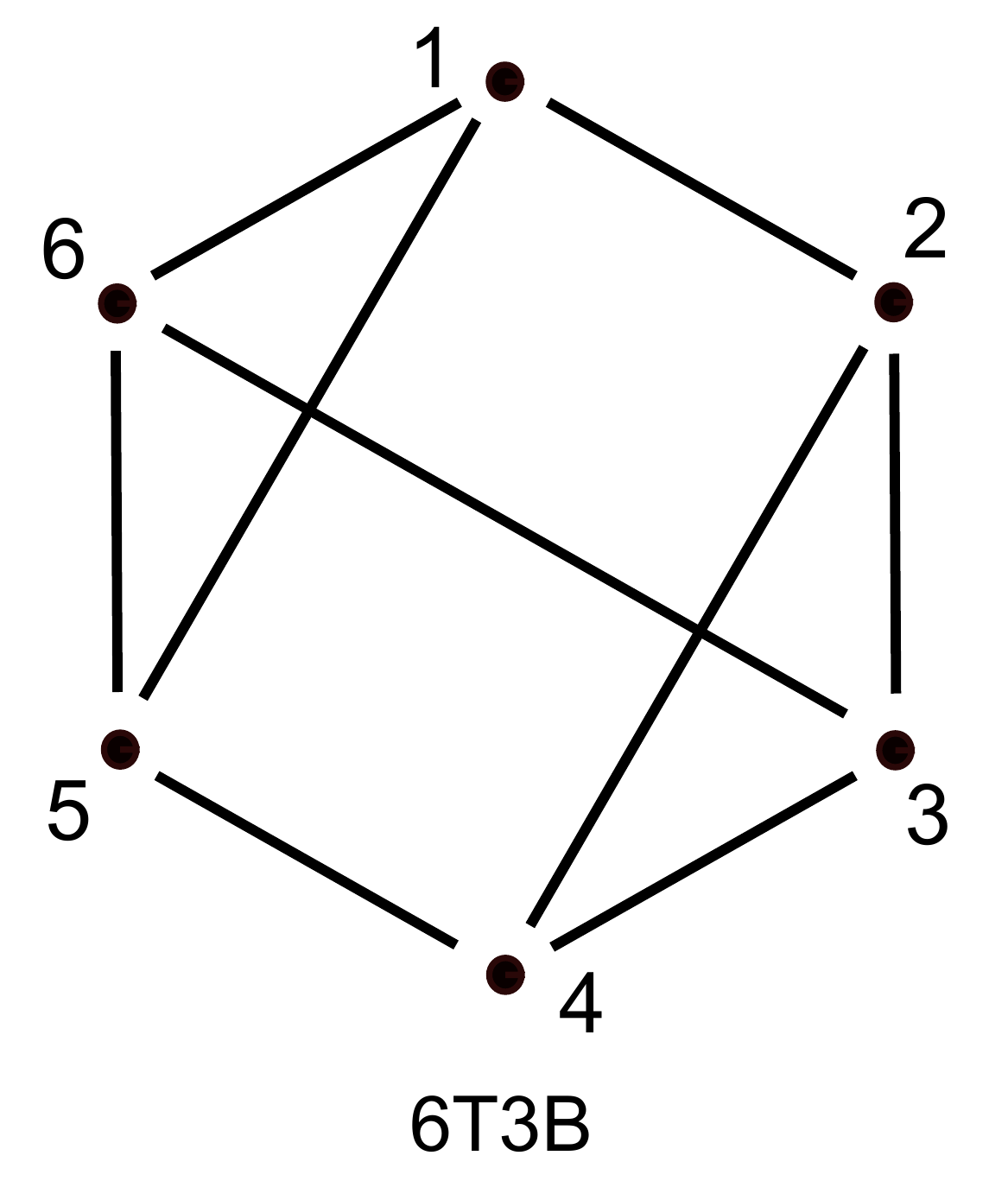}     &
			\includegraphics[width=0.14\textwidth]{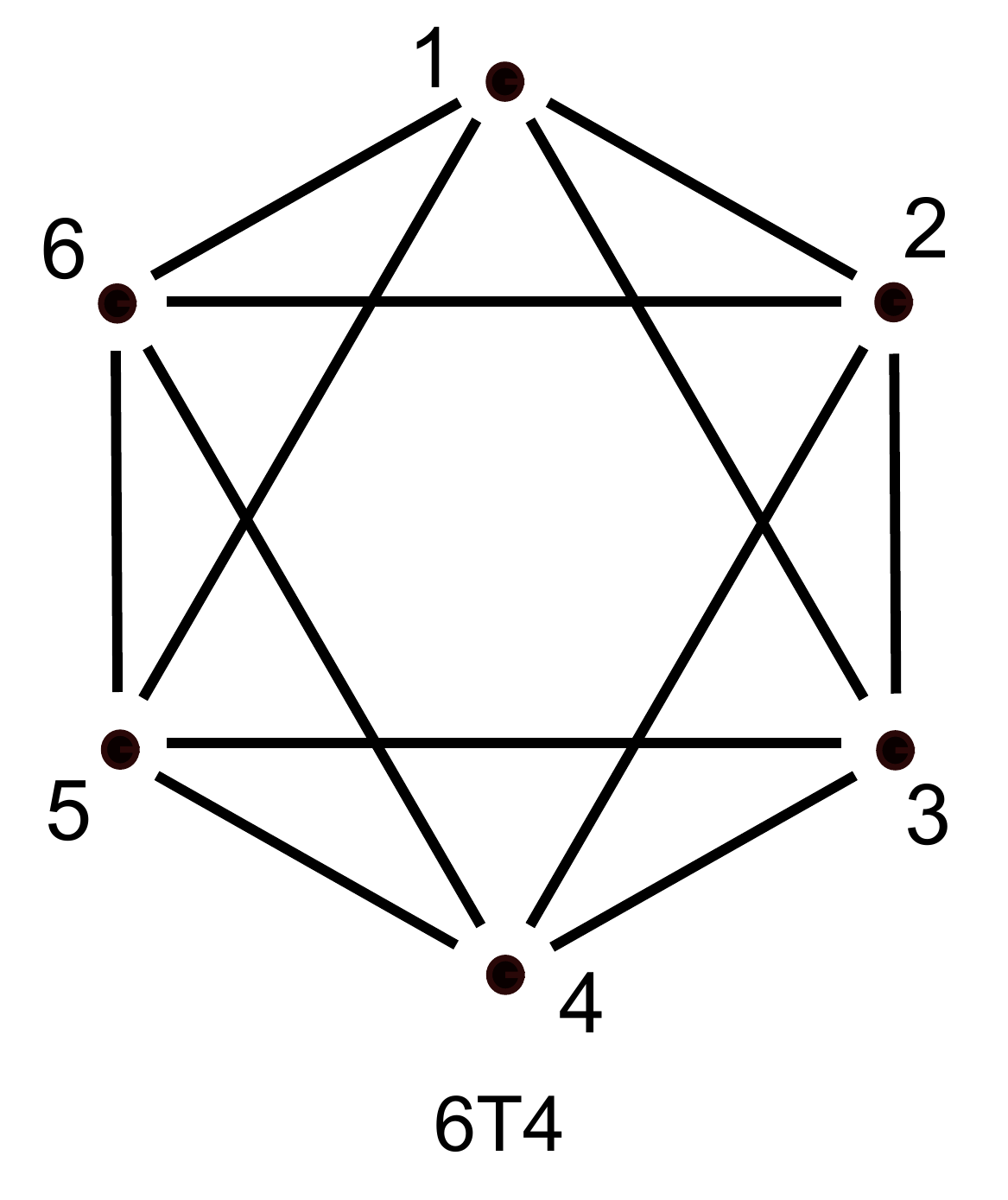}     &
			\includegraphics[width=0.14\textwidth]{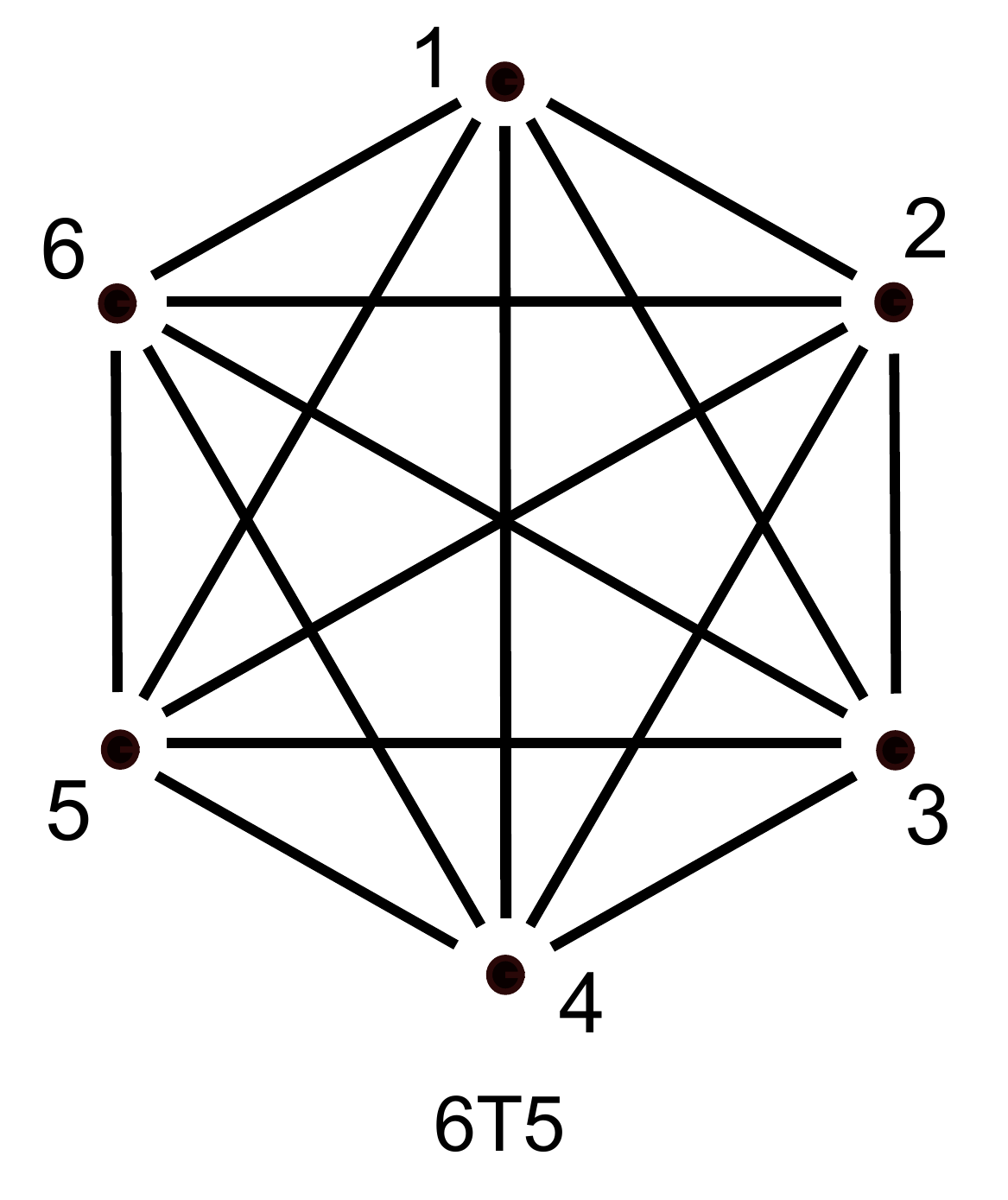}     \\
		\end{tabular}
		\caption{The conceptual combination schemes studied in the 4-telescope (top) and 6-telescope (bottom) cases, with the associated nomenclature below each one.}
		\label{fig:schemes}
\end{figure}

The POPS concept therefore concentrates on co-axial and pairwise solutions, which allow the creation of  the desired number of baselines, thereby limiting the number of pixels to read (to the maximum 24 pixels for a 4T concept with an ABCD coding), and enhancing both sensitivity and speed of the fringe sensor. However when the number of baselines increases each one is less sensitive because the incoming flux has to be divided between more baselines. The sensitivity of the fringe sensor then depends on a competition between the information redundancy between the baselines and their sensitivity. To determine the optimal pairwise co-axial architectures in the 4T and 6T cases, we study the intrinsic performance of several schemes and also point out some important operational aspects (the detailed study is presented in Ref.~\citenum{blind_2010}, in preparation). The schemes studied are presented in the Figure \ref{fig:schemes} with the associated nomenclature.

%
%
%
%

\paragraph{Performance study} The performance study is based on theoretical models of the error on the phase measurement\cite{tatulli_2010}, and is similar to the one led by Houairi et al\cite{houairi_2008} in the context of the GRAVITY fringe tracker. This study consists in computing the optimal optical path estimators used to drive the delay lines, from the noisy phase measurements made on each baselines. When computing the optimal optical paths, the measurements are weighted depending on the measurement error, so that the importance of the noisiest baselines is lowered with respect to the others. 

The quantity of interest is finally the relative error $\epsilon$ on the corrected differential piston which are calculated with respect to a reference noise. The expression of this latter only depends on the considered regime (i.e. detector or photon noise regime), so that results for the different schemes are perfectly comparable within the same regime. More precisely we study three different cases:
\begin{enumerate}
	\item \textbf{Ideal case}: all the baselines are strictly equivalent in term of flux and visibility.
	\item \textbf{Resolved source case}: we assume the source is highly resolved by one baseline (the $\{12\}$ for instance), so that the fringe visibility is set to 0.1 on it, and to 1 on the others.
	\item \textbf{Low coupling efficiency case}: because of quick variations of the instantaneous Strehl ratio, the coupling efficiency on one pupil can be very low compared to the others. To study the behavior of the schemes with respect to this effect, the flux of the telescope $1$ is set to one tenth of the others.
\end{enumerate}
For each case, looking at the relative error $\epsilon$ on the corrected OPD on the different baselines, we are able to compare the combination schemes and to make a choice from a performance point of view. The results of the performance study are on the Table \ref{tab:scheme_perf} for the 4T schemes.

\paragraph{Extracting the photometry} Some schemes also allow to extract the instantaneous photometry on each pupils without the need of dedicated outputs. We show that it is possible only when the schemes have a particular structure: to evaluate the photometry from a pupil, this latter has to be part of a closed sub-array constituted with an odd number of pupils. Otherwise the system linking the fringe signals to the photometries is degenerated. Thus, the 4T2 cannot extract the photometry since it is a cycle of 4 telescopes, whereas the 4T3 can, since there are triangular sub-arrays. The results of this study are presented in Table \ref{tab:photometry}.

%
%
%
%
%
\begin{table}[t!]
	\begin{center}
		\caption{Results of the performance calculations made for the 4T schemes. The relative error is given in detector ($\epsilon^{det}$) and photon ($\epsilon^{phot}$) noise regimes, in the three cases. In the second case, the resolving baseline is the $\{12\}$, and is therefore not equivalent to the others (noted $\{ij\}$). In the third case the baselines affected by the low coupling are noted $\{1j\}$.}
		\label{tab:scheme_perf}
		\begin{tabular}{cccccccccccccccccccc}
		\\
		
\hline
\hline
			   &&& \multicolumn{2}{c}{Ideal case} &&& \multicolumn{5}{c}{Resolved case} &&& \multicolumn{5}{c}{Low coupling efficiency case} \\
 Schemes &&&  $\epsilon_{ij}^{det}$ & $\epsilon_{ij}^{phot}$ &&& $\epsilon_{12}^{det}$ & $\epsilon_{ij}^{det}$ && $\epsilon_{12}^{phot}$&  $\epsilon_{ij}^{phot}$ &&& $\epsilon_{1j}^{det}$ & $\epsilon_{ij}^{det}$ && $\epsilon_{1j}^{phot}$&  $\epsilon_{ij}^{phot}$\\
\hline
4TO &&& 1.6 & 1.3 &&& 16.2 & 1.6 && 13.1 & 1.3 &&& 5.1 & 1.6   && 3.1 & 1.3 \\
4T2 &&&  1.7 & 1.2 &&& 3.4  &  2.0 &&  2.4  &  1.4 &&&  4.7  &  1.9  &&  2.5  &  1.4 \\
4T3 &&&  2.1 & 1.2 &&&  3.0  &  $2.2$ && 1.7  &  $1.3$ &&&  5.6  &  2.4 &&  2.5  &  1.4 \\
\hline
		\end{tabular}
	\end{center}
\end{table}
%
%
\begin{table}[b!]
\centering
\caption{\label{tab:photometry} Capability of the considered schemes to provide the inputs photometries without dedicated outputs.}
\begin{tabular}{cccccccccccc}
\\
 \hline \hline
 Schemes       && 4TO & 4T2 & 4T3 && 6TO & 6T2 & 6T3A & 6T3B & 6T4 & 6T5 \\
 Photometries? && no & no  & yes  && no & no  & no  & yes    & yes  & yes \\ 
 \hline
\multicolumn{10}{c}{}\\
\end{tabular}
\end{table}
%
%

		\subsection{Choice for the 4T case}

\begin{figure}[b!]
	\centering
  	\includegraphics[width=0.49\textwidth]{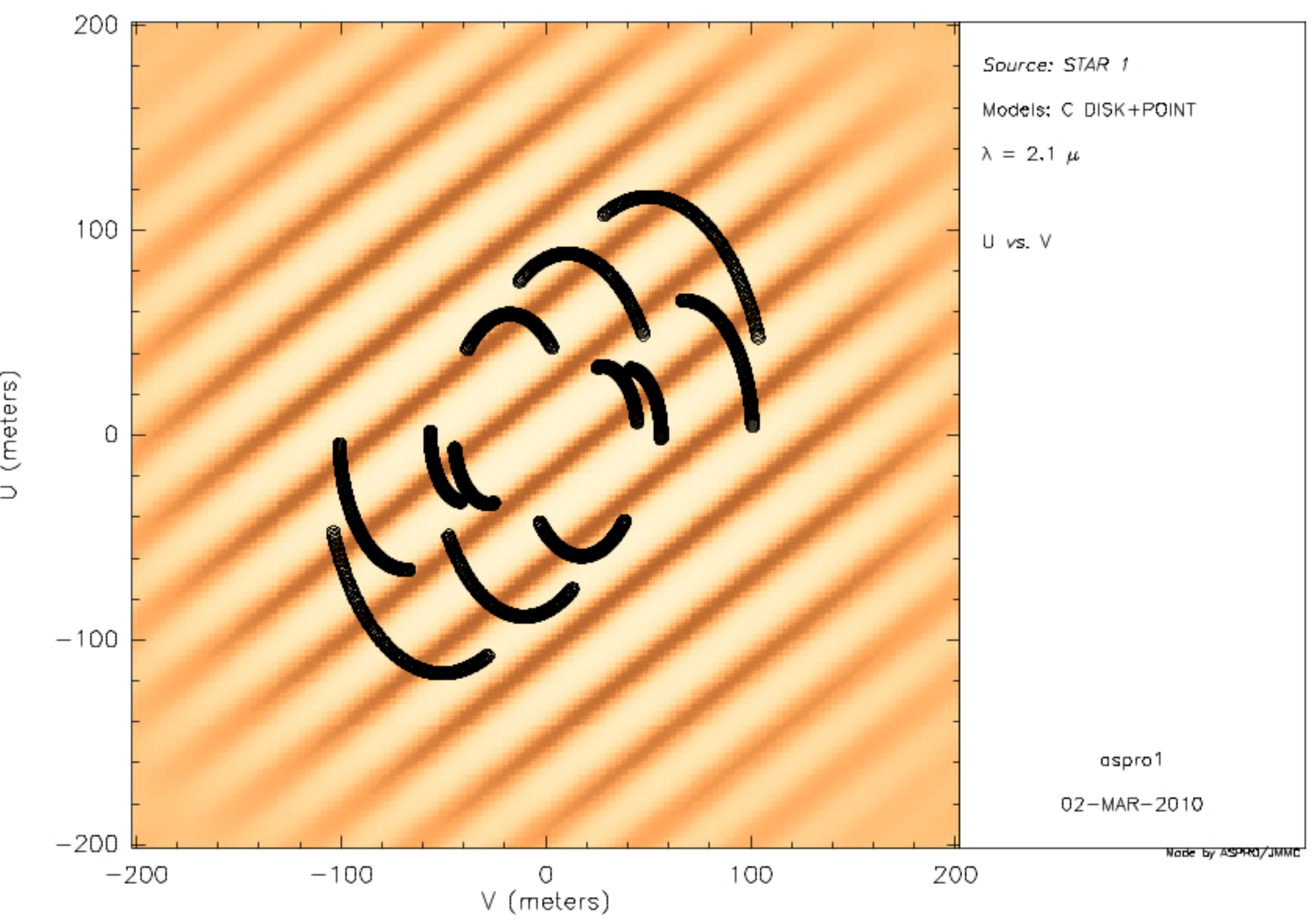}
  	\includegraphics[width=0.49\textwidth]{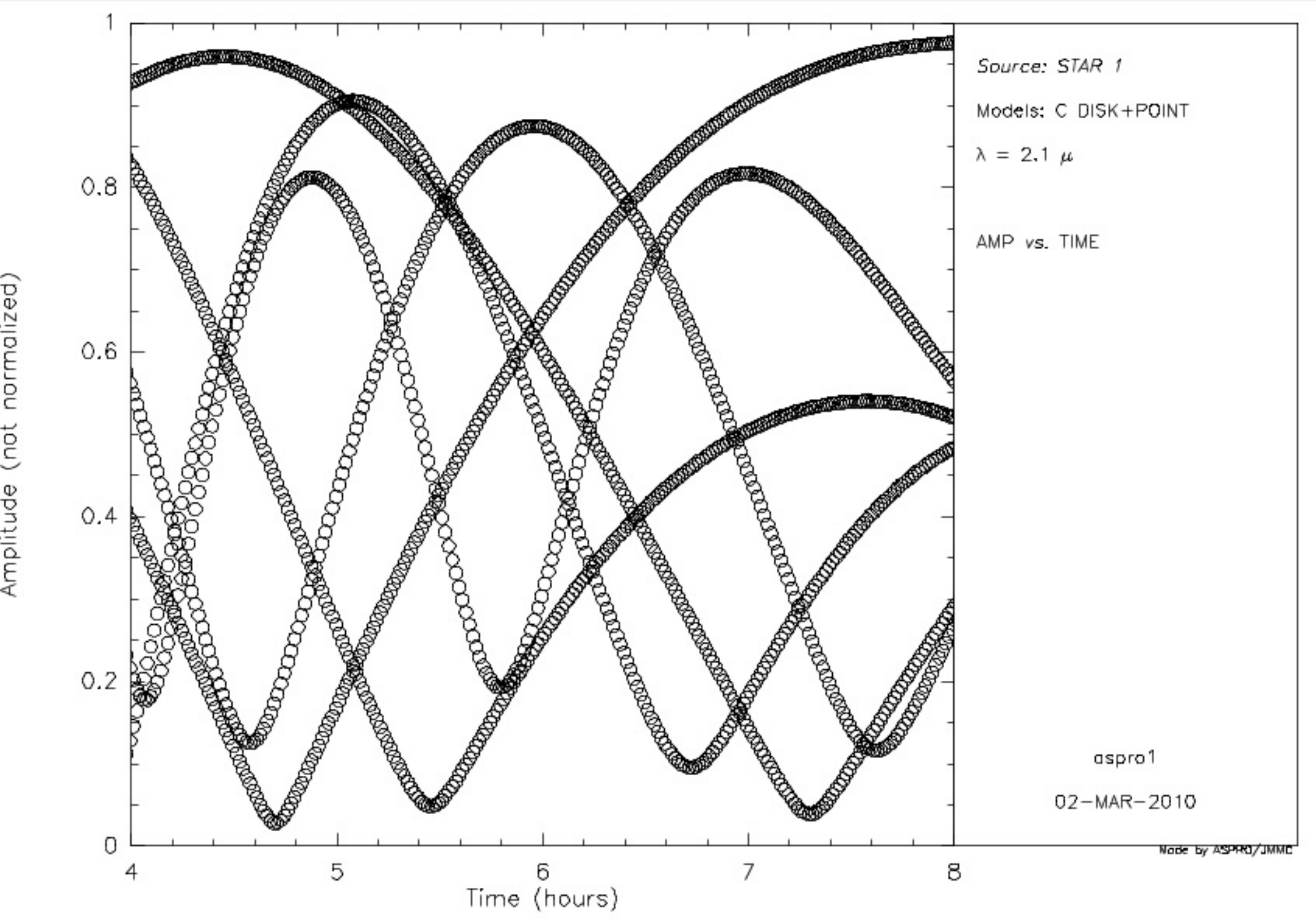}
  	\caption[Predicted fringe contrast when observing a binary]{ Predicted fringe contrast when observing a binary with equal fluxes and a separation of about 10mas with the four UTs and a fringe sensor working in the K band. The left panel shows the (u,v) tracks overlaid on the fringe contrast from the model. Right panel shows the fringe contrast versus time for 4h for each baseline. The figures have been made with the \texttt{aspro} public software from JMMC.}
  	\label{fig:contrast_on_resolved_binary}
\end{figure}

Looking at the performance (Table \ref{tab:scheme_perf}), the open scheme (4TO) appears less interesting than the cyclic one (4T2, with the minimum redundancy) because of its lower redundancy and of the flux imbalance on the extreme baselines, inducing a sub-optimal use of the input photons
\footnote{This observation leads us not to consider schemes with intrinsically imbalanced photometric inputs.}
. The 4T2 and 4T3 appear roughly equivalent: the former provides slightly better performance in the first and third cases, whereas in the second one the 4T3 shows better performance thanks to a more important redundancy. This higher degree of redundancy also provides the following additional advantages, mainly related to operational aspects:
\begin{itemize}
	\item The photometry of each input beam can be directly recovered from the fringe signal (see Table \ref{tab:photometry}). The knowledge of the photometry is theoretically not mandatory to measure the fringe phase, but operation without photometric information is practically possible only for perfectly symmetric combiners. Otherwise, the fringe signal is polluted by photometric fluctuation residuals and the phase and group delay measurements are biased.
Additionally, a real-time photometric monitoring is very useful during operation, providing an additional diagnosis in case of flux drop-outs and allowing the injection to be optimized in all beams simultaneously.
	\item As seen from the VLTI (and so from the sky), all the baselines of a fully redundant pairwise combination scheme are perfectly identical. Whatever the observed target and the geometry of the telescope array, the way to map the interferometer beams into the fringe sensor inputs is always optimal. This is a critical point since during a night unpredictable baselines can exhibit very low visibilities, changing with a time scale of less than one hour (see a typical example in Figure~\ref{fig:contrast_on_resolved_binary}).
	\item Finally, such a strategy also maximizes the cophasing stability in case of frequent flux drop-outs since unaffected baselines are kept locked and thus only the extinguished beam(s) have to be cophased back after extinction.
\end{itemize}
Therefore the fully redundant pairwise combination 4T3 has been chosen. This result is in agreement with the conclusions of Houairi et al\cite{houairi_2008} for the GRAVITY fringe tracker, also based on IO technology.

%
%
%
%

		\subsection{Choice for the 6T case}

The performance study in the 6T case leads to the same kind of conclusion than previously. As for the 4TO scheme, the 6TO open scheme is still penalized because of the imbalance of the inputs and of a lower redundancy. And when the number of baselines exceeds 6, the more important redundancy roughly counterbalances the loss of signal-to-noise ratio on the individual baselines. The 6T3B scheme is finally favored to the 6T2, for the same reasons as in the 4T case (i.e. nearby performance, the possibility to directly extract the photometry from the fringe signal and a higher number of baselines enhancing the robustness).

\section{THE FRINGE SENSING CONCEPTS CHOICES} \label{part:sensing_concepts}

Fringe tracking consists in: \textit{1)} Ensuring the measurements are done in the center of the fringe packet, that is in the highest signal-to-noise ratio (SNR) area, thanks to the group delay measurement; \textit{2)} Keeping the fringes locked thanks to the real-time knowledge of the phase, then allowing long integration times and improved SNR on the scientific instrument. We consider in the following that a fringe tracking regime requires a precision on the phase better than $\lambda/10$, since the contrast loss due to residual fringe blurring is kept below 5\% on the scientific instrument. To fringe track with the highest SNR, it is then mandatory to dispose of phase and group delay estimators as precise as possible, whatever the conditions.
Because POPS is a co-axial combiner based on the IO technology, we consider the ABCD estimator\cite{colavita_1999b}\,, which can be implemented in different ways to measure the phase and the group delay. For  both measurements, we compare  in realistic conditions the performance of two different implementations. These studies are detailled in Ref.~\citenum{blind_2010} (in preparation). We also determine the optimal spectral band and spectral resolution.

	\subsection{Phase sensing} \label{part:phase_sensing}

For the phase we compare two different implementations of the ABCD algorithm: \textit{1)} Four phase states are recorded simultaneously over four different outputs (spatial ABCD); \textit{2)} The phase measurement is made in two steps (temporal ABCD): two phase opposite outputs are first recorded with a zero delay and then with an additional $\lambda/2$ delay. The comparison is based on a theoretical analysis of the signal in both cases, and takes into account atmospheric disturbances (i.e. piston and injection fluctuations). The result in the case of the ATs and an integration time of 1\,ms is presented in Figure \ref{fig:snr_phase}.

This study shows that the non-simultaneity of the measurements in the temporal concept introduces an important, additional noise, mainly due to piston. Because of it, the temporal ABCD is always less efficient in a phase tracking regime, so that the spatial ABCD appears as the best choice. Especially, in bad seeing conditions or on UTs, the phase error can be higher than $\lambda/10$, limiting the phase tracking capabilities.

	\subsection{Group delay sensing} \label{part:group_sensing}

Still in the context of the ABCD coding, we consider two concepts to measure the group delay : \textit{1)} The scanned-fringes concept where the fringe contrast is measured on some points inside the main lobe of the fringe packet; \textit{2)} The dispersed-fringes concept, where the fringe phase is measured over a few spectral channels.

\begin{figure*}[t]
	\begin{minipage}[t]{0.49\linewidth}
		\centering
		\includegraphics[width=0.8\linewidth]{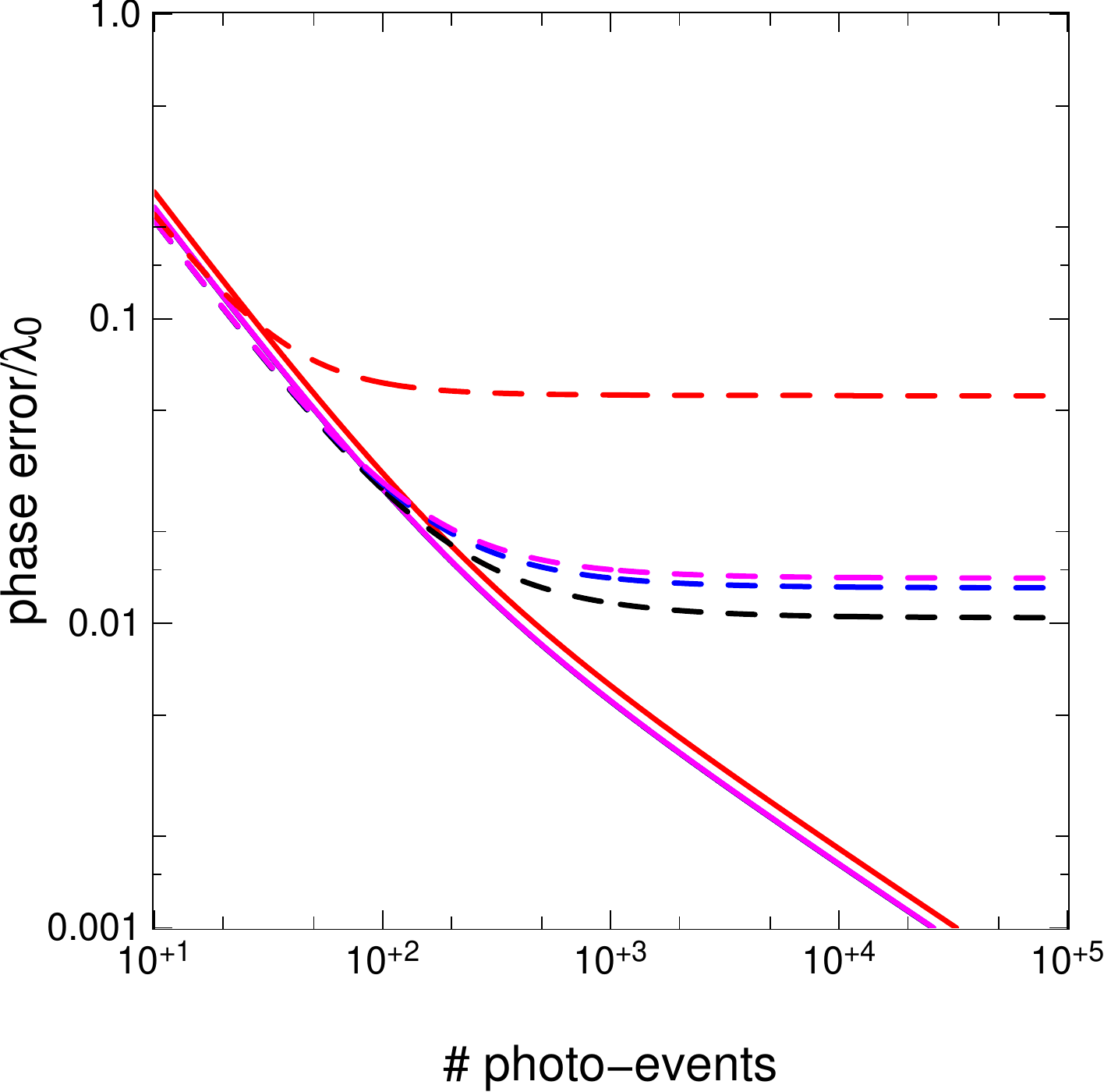}
		\caption{Phase measurement statistical error functions of the incoming flux in the ATs case with an integration time of 1\,ms, for the spatial ABCD (solid) and temporal ABCD (dashed) phase sensing concepts. From black to red, the atmopsheric conditions vary from excellent to bad.}
		\label{fig:snr_phase}
	\end{minipage}
	\hfill
	\begin{minipage}[t]{0.49\linewidth}
		\centering
		\includegraphics[width=0.8\linewidth]{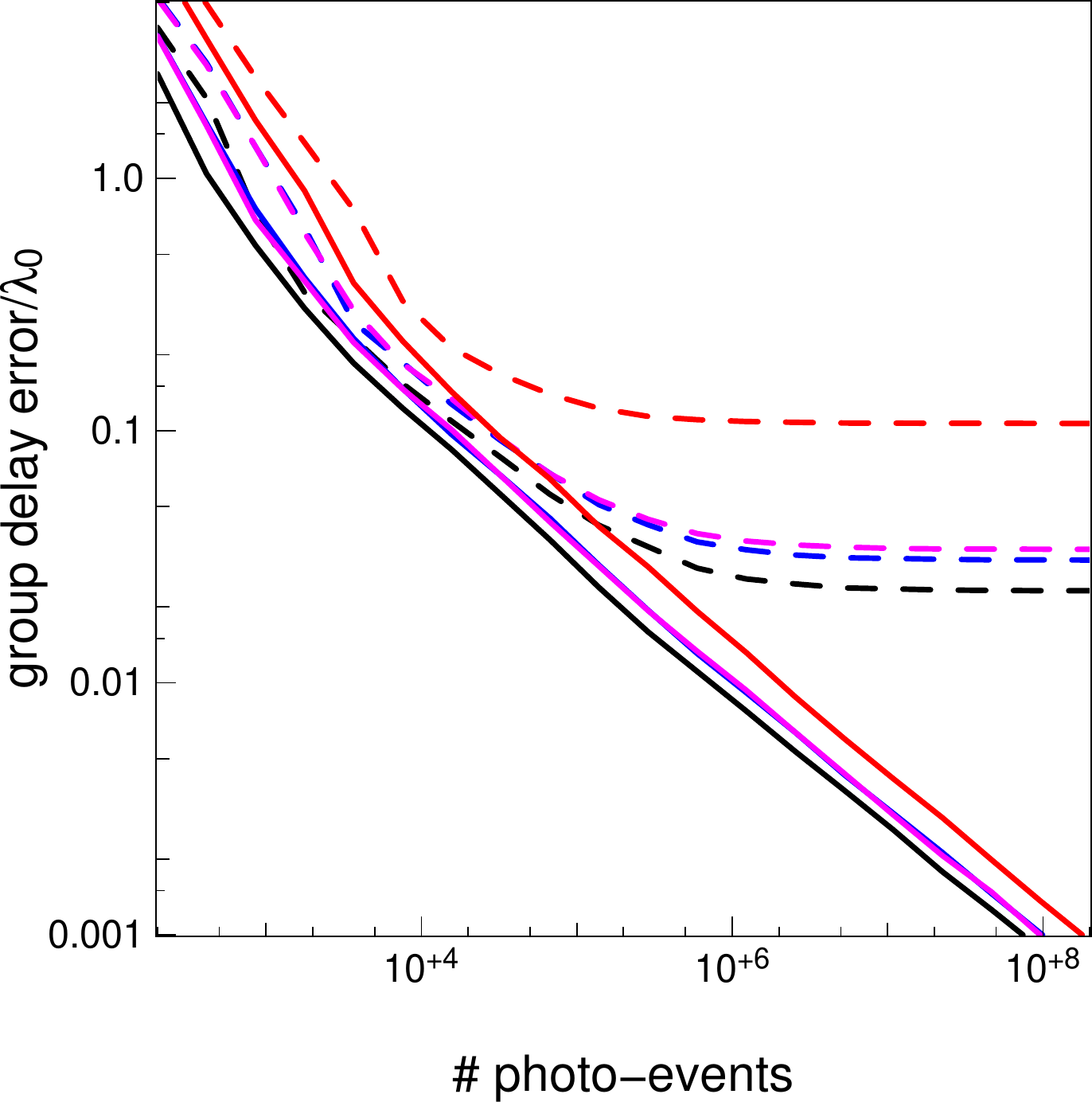}
		\caption{Statistical error for the group delay estimation with the dispersed-fringes (solid) and the scanned-fringes (dashed) group sensing concepts. Curves are given functions of the incoming flux in the ATs case for an integration time of 1\,ms. From black to red, the atmopsheric conditions vary from excellent to bad.}
		\label{fig:snr_gd}
	\end{minipage}
\end{figure*}

Given the impossibility to carry a realistic analytical description of these group delay estimators, we proceed with Monte-Carlo simulations taking into account atmospheric disturbances. The comparison of the dispersed-fringes and the scanned-fringes concepts, in various atmospheric conditions, shows the first is the most efficient (the results of simulations in the case of the ATs with an integration time of 1\,ms are presented on the Figure \ref{fig:snr_gd}). Equally as the temporal ABCD for the phase, the scanned-fringes concept exhibits a very high sensitivity to atmospheric disturbances (piston mainly) so that the dispersed-fringes concept appears as the best choice. For instance, even for integration times as low as 0.5\,ms, the precision of the temporal concept is worse than $\lambda/10$ for the ATs in bad atmospheric conditions and for the UTs whatever the conditions.

%
%
%
%

	\subsection{Spectral disperion} \label{part:spectral_disp}

The number of spectral channels directly determines the efficiency of the dispersed group delay sensing. A trade-off should be made between an accurate determination (larger spectral dispersion) and an optimal sensitivity when limited by detector noise (smaller dispersion). We decide to disperse the light over 5 channels, corresponding to a spectral resolution $\sim 20$ in order to provide a non-ambiguous range of $\pm 20 \mu m$ in which the group delay can be found. This non-ambiguous range is larger than the OPD typically introduced by the turbulence over few seconds ($\approx 10\mu m$ at Paranal). Therefore the group delay can be quickly recovered after a temporary flux loss (low Strehl or telescope chopping for MATISSE). This non-ambiguous range is also significantly larger than the size of the fringe packet seen in broad band. The detection of the main lobe of the fringe packet is much easier, and thus avoids locking the fringes on a secondary lobe, with a low SNR.

%

	\subsection{Fringe sensing in the K band} \label{part:Kband}

All required optical devices (IO, fibers, bulk) exist in both the H and K astronomical bands. We used the simulation tool \simPOPS{} (see section \ref{part:sim2GFT} for details) to compute the expected performance in both bands, taking into account the differences in atmospheric turbulence and transmission, sky brightness and expected instrumental transmission.

We finally choose the K band from $1.9$ to $2.4\,\mu m$ because it presents a higher limiting magnitude ($\sim 0.3$\,mag). The effective wavelength is also 35\% longer enhancing the non-ambiguous phase delay and group delay measurement range, and the atmospheric turbulence strength is smaller, leading to a higher and more stable flux injection in the optical fibers. Furthermore, longitudinal and transversal atmospheric dispersion compensation (LADC and TADC) is not required considering the low losses that are expected (typical coupling losses of about $20\%$ and typical fringe contrast loss of $10\%$ for a zenith distance Z$=60\deg$).


On the other hand, the K band suffers from an increased thermal background with respect to the H band. The IO component has to be kept cooler than about $240$K to limit its emission in the $1.9-2.4\mu m$ spectral range.

%
%
%
%
%
%
\begin{figure}[tb]
	\centering
		\includegraphics[width=.5\textwidth]{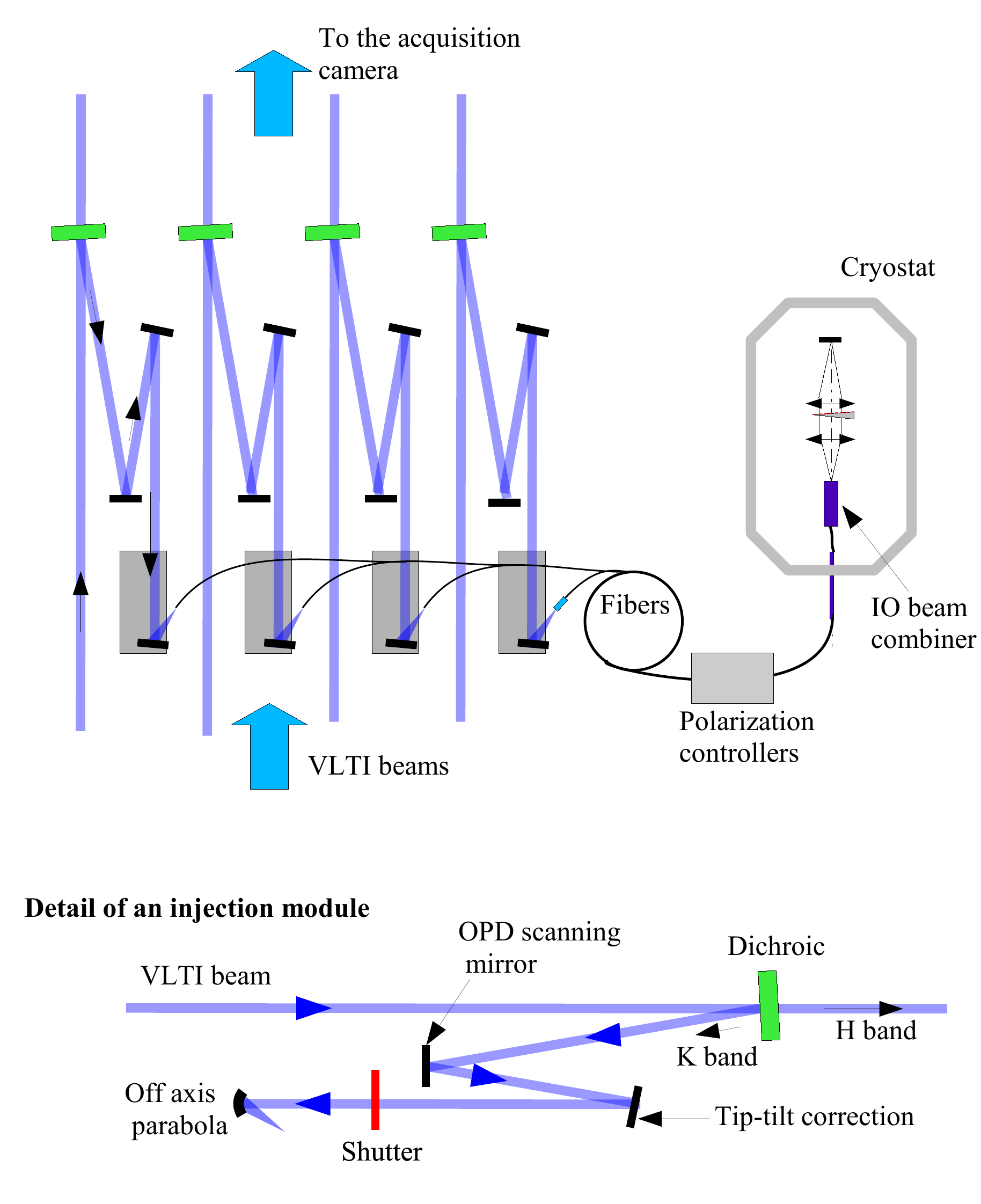}
	\caption[Schematic view of the POPS concept]{\label{fig:concept_scheme} Schematic view of the 4T POPS concept (top) with the 4 modules injecting the VLTI beams into single-mode fibers, and combining them into the IO component. This latter is placed with the camera inside the cryogenic dewar. At the bottom is a detailled view of an injection module.}
	\label{fig:concept_design}
\end{figure}
%
%

\section{PROPOSED CONCEPT} \label{part:proposed_concept}

Based on the conclusions of the previous sections, the 4-telescope POPS concept has finally the following characteristics: \textit{a)} The beam combination is performed thanks to a single-mode integrated optics component using a fully redundant 4T3 scheme with internal spatial ABCD coding; \textit{b)} The phase is estimated thanks to the ABCD outputs, and the group delay is measured by additionaly dispersing them over 5 channels; \textit{c)} The concept is optimized for the K spectral range ($1.9$ to $2.4 \mu m$) and needs no TADC nor LADC.

	\subsection{Concept overview} \label{part:overview}

Figure \ref{fig:concept_design} presents a schematic view of the overall concept as well as one of the injection module in the bottom part. The VLTI beams are collected through 4 injection modules, insuring several functions, before to be injected into the SM fibers and the IO beam combiner. The outputs of the IO chip are imaged and dispersed over 5 pixels of the Hawaii II RG camera. Since the fringe coding is based on an internal ABCD modulation, there is no particular need for an external metrology.

With this design, the transmission of POPS is estimated to roughly 23\%, and the instrumental contrast to 87\%. The concept footprint is 1000$\times$2500\,mm in the 6T configuration.

	\subsection{Injection unit} \label{part:injection}

The injection assembly is based on the PIONIER\cite{berger_2010} visitor instrument that is currently under integration at LAOG and provides the following functions (see Figure \ref{fig:concept_design}, bottom):

\vspace{0.1cm}
\noindent \textbf{Selection of the K band} \quad The VLTI beams are collected by dichroics (in green) reflecting the K band in POPS while the H band is transmitted to the IRIS guiding camera.

\vspace{0.1cm}
\noindent \textbf{Tip-tilt correction} \quad A tip-tilt mirror provides a real-time compensation of the tip-tilt variations created in the VLTI tunnels.

\vspace{0.1cm}
\noindent \textbf{Calibration} \quad An optical path modulator and a shutter are used for internal calibration of the beam combiner (instrumental contrast, ABCD phase shifts, etc.). We use a long-stroke modulation in order to have a sufficient spectral resolution.

\vspace{0.1cm}
\noindent \textbf{Injection} \quad The VLTI beams are coupled into a SM fiber through an off-axis parabola. The fiber and the parabola are mounted on a common support, so that the image quality on the fiber head can be adjusted conveniently. The mechanical support of the injection module embeds a manual OPD adjustment to cophase all arms.

%
%
\begin{figure}[b]
	\centering
	\includegraphics[width=0.65\textwidth]{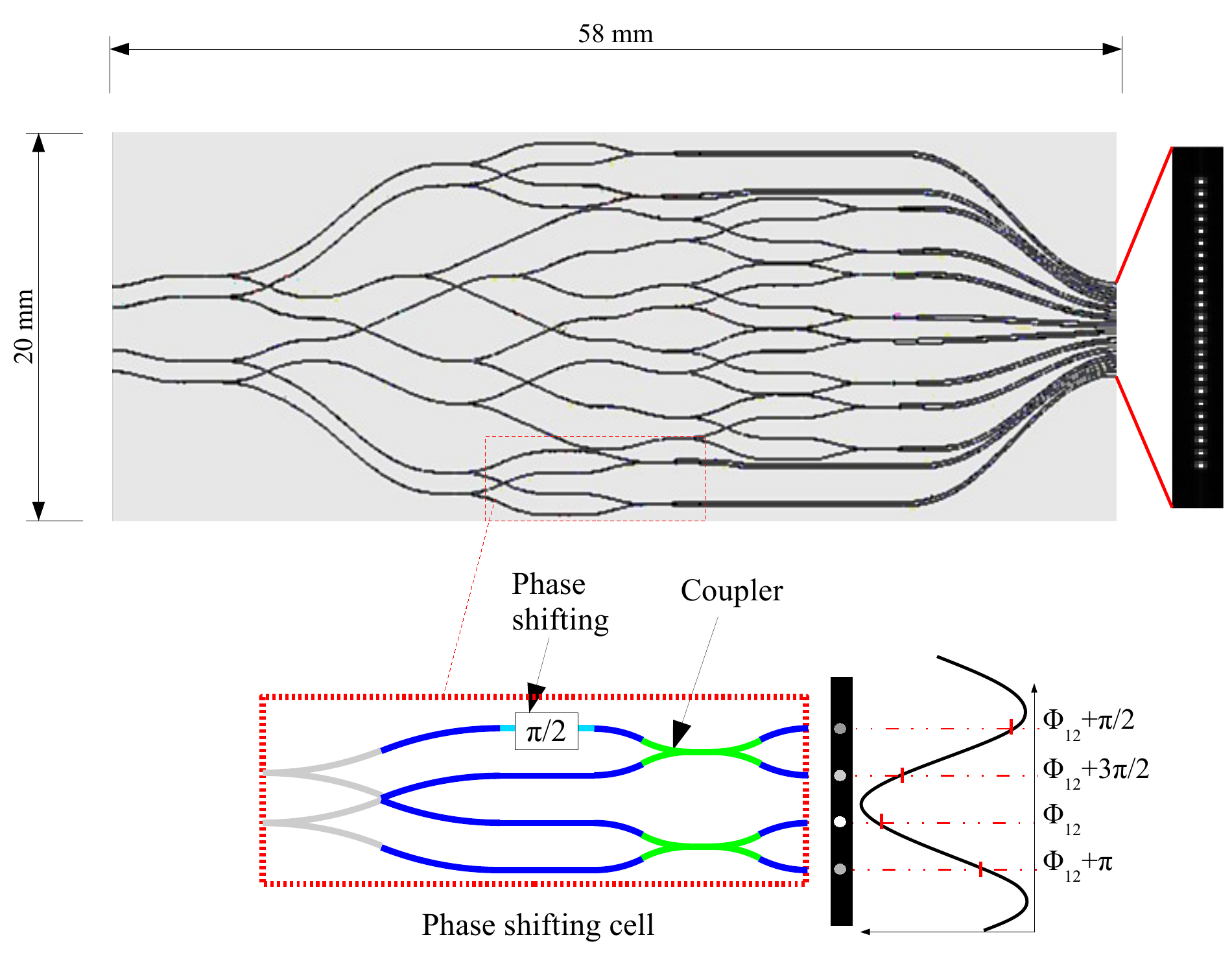}
	\caption[Design of the 4T3-ABCD GRAVITY beam combiner. ]{Design of the 4T3-ABCD beam combiner. The 24 interferometric outputs that are imaged on the camera are on the right hand side of the figure.}
	\label{fig:IO-4TR3}
\end{figure}
%
%

%
%
\begin{table}[t]
	\begin{center}
	\caption{Expected performance of the 4T3 IO component in K band.}
	\label{tab:IOperf}		\begin{tabular}{ll}
	\\
		\hline
		Average transmission  & $>55\%$ \\
		Broadband contrast level  &	$> 90 \% \pm 5\%$, with a Wollaston prism.\\
 		Narrowband contrast level &  $95\%$, with a Wollaston prism. \\
 		Phase shifting value & $90\deg\pm25\deg$ \\
 		Phase shift Chromaticity & $5$ to $15\deg$ \\
		Phase shift between polarization & $5\deg$ to $10\deg$ \\
 		Contrast stability & $\leq 0.5 \%$ \\
		Phase shift stability & better than $1\deg$ \\
		\hline
\\
		\end{tabular}
	\end{center}
\end{table}
%
%

	\subsection{Beam combination unit} \label{part:IOC}

This assembly is composed of the integrated beam combiner and its feeding fibers. Working in the K band, the design of this assembly is based on the one currently developped for GRAVITY. It is consequently based on fluoride fibers to maximize the throughput of POPS for K band operation, especially for wavelengths above $2.3\,\mu m$. They are mounted in a fiber array V-groove glued to the IO component. This assembly is then integrated in the cryostat in order to insure the required stability between the component output and the detector itself. It allows also cooling down the component and the imaging optics at 240K to achieve a negligible background emission.

\vspace{0.1cm}
\noindent \textbf{IO combiner} The concept chosen for POPS is the 4-beam pairwise ABCD combining 6 baselines (4T3 scheme), resulting in 24 interferometric outputs. The circuit, designed in collaboration with our partner CEA/LETI, is presented in Figure~\ref{fig:IO-4TR3}. The combiner includes a series of achromatic phase shifters to produce four phase states in quadrature, resulting in a spatial ABCD coding for each baseline.

A prototype in the H band has been fully characterized by Benisty et al\cite{benisty_2009} . The equivalent circuit for the K band is currently under manufacturing for GRAVITY and the expected performance are evaluated from a set of prototypes tested in laboratory (Table~\ref{tab:IOperf}). The current spectral transmission in K band also presents a strong OH absorption peak at $2.2\, \mu m$, so that the average transmission over the K band should be $>55\%$.

\vspace{0.1cm}
\noindent \textbf{Polarization control}	Using low birefringence fibers, polarization control is required to optimize the contrast level. Each fiber arm would therefore implement a fibered polarization controller motorized with a stepper motor, similar to those designed for GRAVITY.

	\subsection{Cryogenic unit} \label{part:cryo}

The cryogenic mechanical structure does not include any manual or motorized adjustment to ensure a high stability level. The cryogenic dewar is based on the last generation of cryostats developped by ESO. It cools down the beam combination optics, the imaging optics and the Hawaii II RG camera. Since POPS is designed for operation with 4 or 6 telesopes, two IO components (4T3 and 6T3B) have to be implemented, mounted upside-down on a bracket support in order to have the two outputs slits imaged on the detector as close as possible.


\section{EXPECTED PERFORMANCE OF THE 4T CONCEPT} \label{part:expected_perf}

This final section presents the expected performance of the POPS proposition in the 4-telescope case. We first describe our end-to-end simulation tool \simPOPS{} and then present the computed performance for the ATs, the UTs and the specific case of mixed arrays ATs+UTs.

	\subsection{\simPOPS} \label{part:sim2GFT}

In order to assess the performance of our conceptual designs, and to compare the performance of various designs (i.e., combination of a hardware solution and a phase estimator) under various assumptions in terms of environmental and observing conditions, we have developed a dedicated software simulation tool called \simPOPS{}.

\simPOPS{} is largely based on the \texttt{GENIEsim} software, which is described in detail by Absil et al.\cite{absil_2006} , and therefore follows the same architecture and philosophy. The simulations are taking into account all major contributors to the final performance, from the atmosphere and the telescopes down to the fringe sensor and delay lines.
In order to properly estimate the amount of coherent and incoherent photons, all the VLTI and POPS subsystems are described by their influence on the intensity, piston, and wavefront quality of the light beams collected by each telescope. The implementation of each subsystem in the simulator has been extensively validated during the GENIE study\cite{Absil_master} .

Some additional physical phenomena (visibility loss due to piston jitter, atmospheric refraction, intensity mismatch between the beams due to atmospheric turbulence, and longitudinal dispersion in the delay lines) have been implemented in the fringe tracking loop of \simPOPS{} with respect to what was implemented in \texttt{GENIEsim}. 

%
%
\begin{figure*}[t]
	\begin{minipage}[t]{0.49\linewidth}
 		\centering
 		\includegraphics[viewport=0 10 510 370,clip,width=0.95\textwidth]{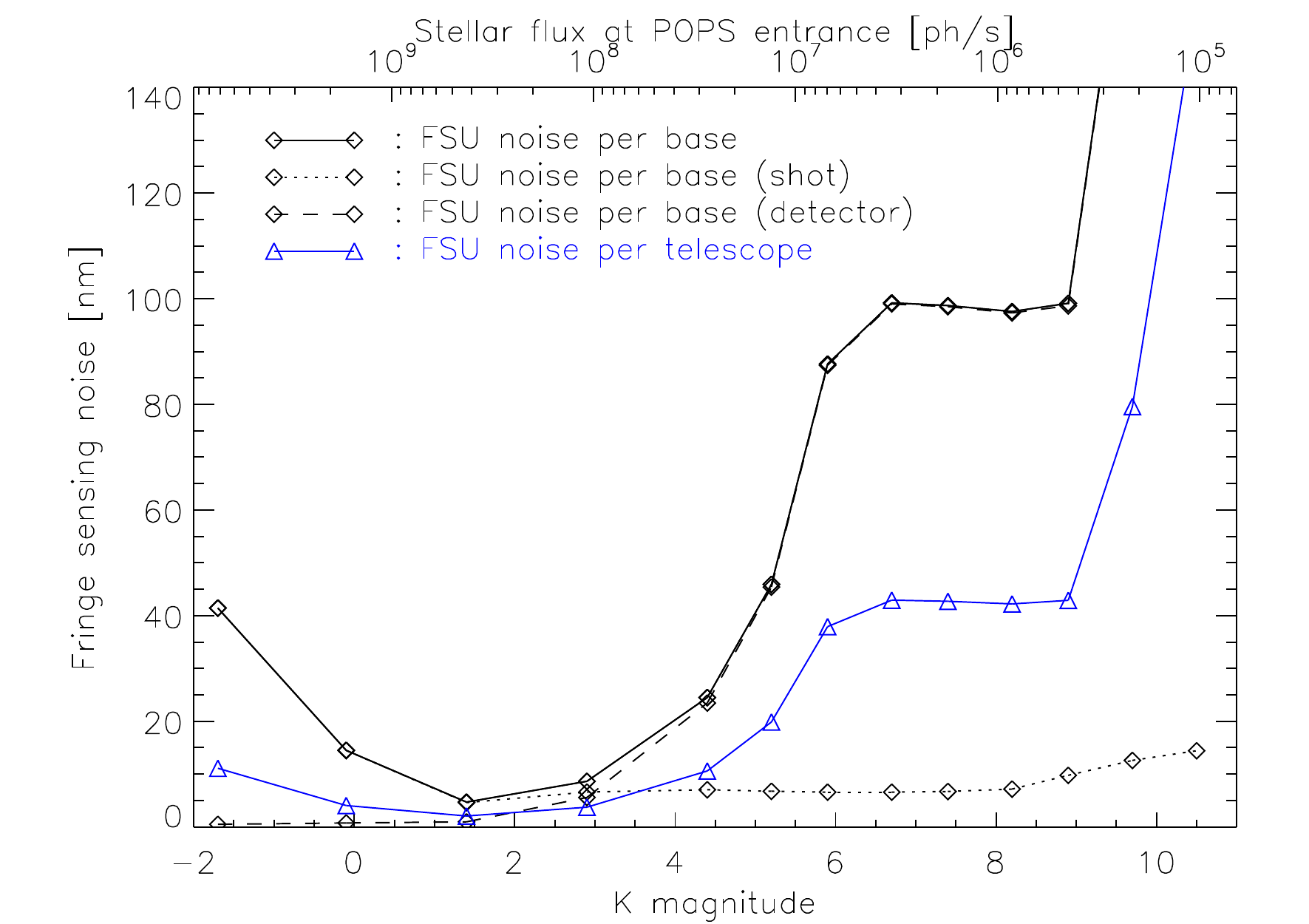}
		\caption[Fringe sensing noise as a function of the target magnitude]{Fringe sensing noise functions of the K-magnitude of the K0\,III star source in the 4T3-ABCD case. The corresponding stellar flux in photons per second at the entrance of POPS is on the top scale. The respective contributions of shot and detector noises are represented by dotted and dashed lines.}
		\label{fig:fsunoise_4TABCD}
	\end{minipage}
\hfill
	\begin{minipage}[t]{0.49\linewidth}
   	\centering
  		\includegraphics[viewport=5 10 500 370,clip,width=0.95\textwidth]{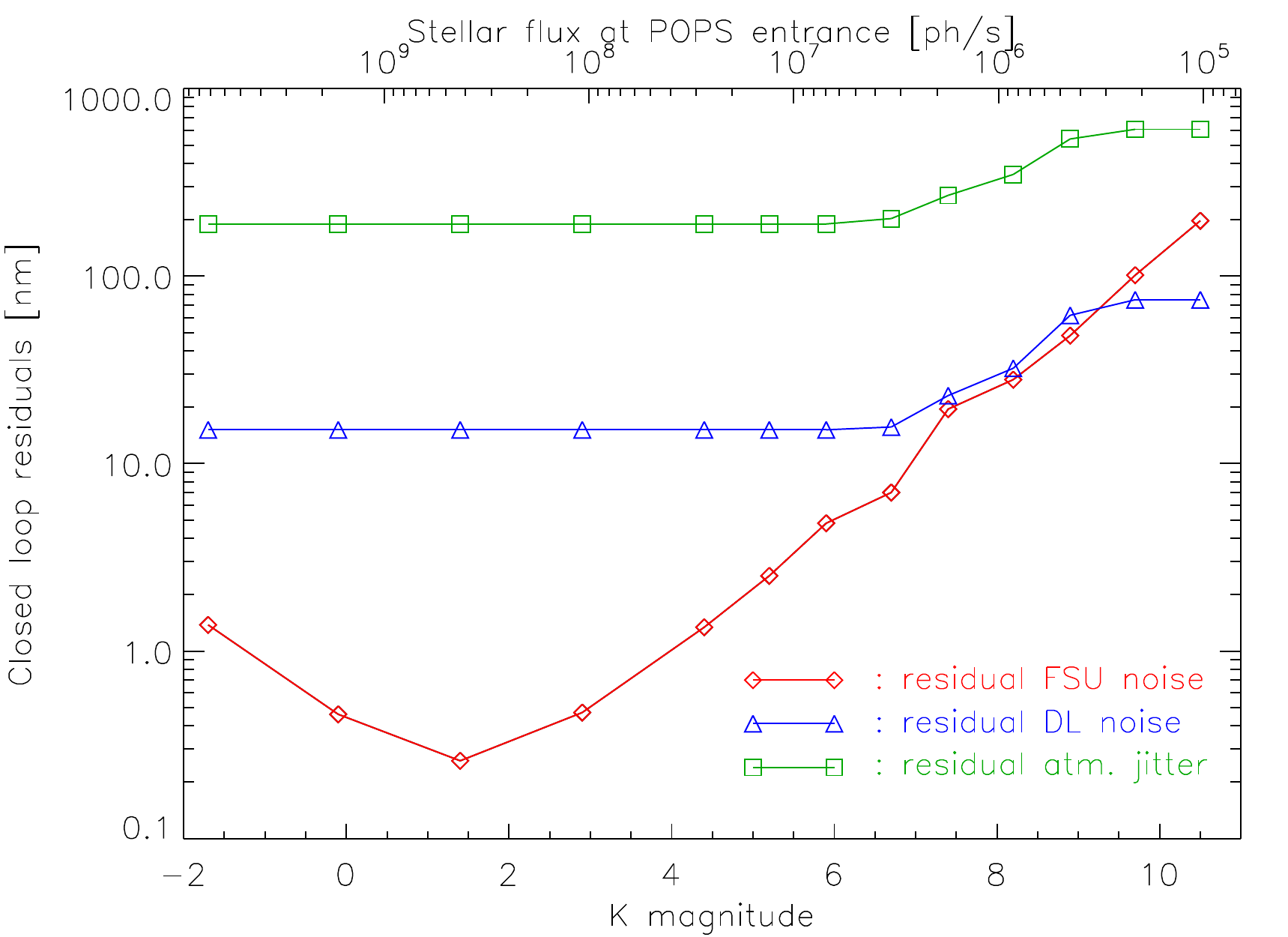}
  		\caption[Repetition time, time delay and noise residuals as a function of the magnitude]{Noise residuals at the output of the closed-loop for the fringe sensing, the delay line and the corrected atmospheric piston.}
  		\label{fig:closeloop_4TABCD}
 \end{minipage}
\end{figure*}
%
%

	\subsection{ATs case}

In the case of the ATs, simulations have been done for 4 different concepts (4T2 and 4T3 schemes associated with a spatial or a temporal ABCD for the phase estimation). They show good agreement with the previous studies, that is the 4T3 with a spatial ABCD concept is the most efficient in various atmospheric or observationnal conditions. They particularly highlight the contribution of the redundancy, providing a more stable fringe tracking: in the case of a resolved binary (see Figure \ref{fig:contrast_on_resolved_binary}), the fully redundant scheme (4T3) outperforms the cyclic one (4T2).

Therefore we present here the results of simulation for the selected concept that is the 4T3 scheme with a spatial ABCD coding of the fringes. The fringe sensing and closed-loop control performance are summarized in Figures~\ref{fig:fsunoise_4TABCD} and \ref{fig:closeloop_4TABCD}, for a range of magnitudes from $K=-1.7$ to $K=10.5$. These simulations have been obtained for a K0\,III star located at various distances ranging from 12\,pc to 3200\,pc, with the A0-G1-K0-I1 quadruplet of ATs, and in standard atmospheric conditions: seeing $\varepsilon=0.85"$, Fried parameter $r_0=12$\,cm, coherence time $\tau_0=3$\,ms, average wind speed of 12\,m/s, outer scale ${\cal L}_{\rm out}=25$\,m, and sky temperature $T_{\rm sky}=285$\,K. The target star is assumed to be located close to zenith.

Figure~\ref{fig:fsunoise_4TABCD} shows the fringe sensing noise (due to shot and detector noises in the fringe sensor) functions of the magnitudeÒ. 
%
On the bright-side end of the plot, shot noise dominates the noise budget. The increase in shot noise from $K=1.5$ to $K=-2$ is due to the star being (strongly) resolved, which reduces the available coherent flux. Detector noise becomes larger than shot noise around $K=3$, and the fringe sensing noise reaches its allowed limit (100\,nm RMS) around $K=6$ . In this regime the loop can be operated at its maximum repetition frequency  of 3559\,Hz. For fainter magnitudes, \simPOPS\ makes sure that the fringe sensing noise remains at the same level by gradually increasing the  closed-loop repetition frequency (i.e., increasing the integration time on the detector) to keep a sufficient SNR on each individual fringe measurement. This is possible only until magnitude $K\sim9$ in the present case, where a phase sensing noise of 100\,nm per baseline cannot be reached any more for any integration time, because of the strong fringe blurring that appears at long integration times. Above this limit, the integration time is adjusted to minimize the measurement noise. Also represented in Figure~\ref{fig:fsunoise_4TABCD} is the fringe sensing noise per telescope, which results from the optimized use of the differential piston measured on all the baselines: it is significantly smaller than these latters thanks to the information redundancy.

Finally, for an integration time of 10\,ms, the limit above which the sensing noise cannot be maintained below 100\,nm is $K=8.2$. If we consider the {\it operational limiting magnitude} for stable fringe tracking to be the limit for which the SNR is larger than $\approx 4$ for at least $95\%$ of the time, we obtain a value $K = 7.2$ for the ATs in standard conditions. The limit for fringe detection (SNR$\approx 4$) is $K=9.2$ in excellent conditions.
 
The noise residuals at the output of the closed-loop are represented in Figure \ref{fig:closeloop_4TABCD}. It shows that the fringe sensing noise is always dominated by the residual atmospheric noise, which is due to the delay between the phase estimation and the moment the correction is applied to the delay lines as well as the finite acquisition frequency of the loop. The fringe sensing noise has however a strong influence on the fringe tracking residuals, as it sets the maximum loop repetition frequency that can be used. In particular, for magnitudes $K>6$, in order to keep a sufficient SNR on the measurements, the loop frequency is reduced, which results in increased closed-loop residuals.

	\subsection{UTs case}

The same performance study has been carried out in the case of the UTs, showing a similar general behaviour as in the case of ATs. The main difference is that the {\it operational limiting magnitude} reaches $K=9.2$, and the limit for fringe detection $K=11.2$. Furthermore, the coupling efficiency decreases for stars fainter than $V=10$, because of the reduced performance of the MACAO adaptive optics system. According to our simulations, one can consider that for V magnitudes higher than $\sim12$, the MACAO correction becomes insufficient to feed the fringe sensor with a beam stable enough. However telescope vibrations are expected to strongly affect the residual piston jitter at the output of the closed-loop (an effect not simulated in \simPOPS).

These expected limiting magnitude are in agreement with the results of instruments in operation. Indeed, PRIMA FSU detected but did not track noisy fringes on a $K=11.7$ target\cite{sahlmann_2009} on the UTs.

	\subsection{Case of mixed array ATs + UTs}

Combining UTs and ATs allows to access to more apertures and to increase the SNR. From a technical point of view, mixing ATs with UTs is possible since the VLTI infrastructure is designed to allow the combination of apertures of different sizes. This case has not been simulated, but simple calculations on the fringe SNR show an increase in the limiting magnitude up to 1.2 mag compared to the ATs+ATs case.

	\subsection{Influence of the atmospheric conditions}
	
The influence of atmospheric conditions on the fringe tracking performance is mainly twofold: on the one hand it determines the input atmospheric noise that needs to be corrected, and on the other hand it affects the amount of available coherent photons since it determines the injection efficiency into SM fibers. The POPS performance are simulated from bad atmospheric conditions (seeing $\epsilon_0=1.1"$ and coherence time $\tau_0=2$\,ms) to excellent ones ($\epsilon_0=0.5"$, $\tau_0=10$\,ms) for the ATs. The main conclusion is that the limiting magnitude increases by about 1~magnitude between the bad and the excellent conditions. With integration time of $10$\,ms the limit to  maintain the sensing noise below $100$\, nm varies between $K=7.7$ and $K=8.8$ depending on the conditions.

\section{CONCLUSION}
POPS is a K band fringe sensor based on the integrated optics technologies. It measures the fringe position (phase and group delay) thanks to a spatial ABCD fringe coding, dispersed over 5 spectral channels. For the moment only the 4T concept is clearly defined and will recombine the 6 possible baselines to improve the fringe tracking stability in varying observationnal and atmospherical conditions. Realistic simulations show the expected performance and limiting magnitude of POPS are similar to those obtained with the current instruments in operation, in particular the PRIMA FSU.

POPS will highly benefit from the technological breakthroughs in the coming years. In particular, the developments of the GRAVITY beam combiners will be directly applicable to the proposed POPS design, especially the improvement of the components transmission and the validation of the polarization controllers.

POPS has also an instrument layout similar to the PIONIER visiting instrument, which will be commissioned at VLTI at the end of 2010. POPS should then benefit from the on-sky experience feedback of PIONIER.

Finally, the availability of fast new generation detectors with smaller read-out noise will impact several of the system choices of POPS. In particular, with such detectors, the multi-axial scheme is significantly less penalized by the number of pixels it requires, and becomes more attractive. Indeed, the multi-axial IO components are shorter and simpler, i.e. more transmissive and probably less chromatic. Furthermore a component designed for 6 telescopes can be used also to combine 4 telescopes, with a limited loss of SNR with respect to a design built for 4T only. Consequently multi-axial combination could be a great alternative with these future detectors. The final choice for the 6T combination concept is therefore still open.

\acknowledgments
\noindent This phase A study was supported by funds provided by ESO and has made use of the Jean-Marie Mariotti Center \texttt{Aspro} service \footnote{Available at http://www.jmmc.fr/aspro}.

%
%

\bibliographystyle{spiebib}
\bibliography{bibPOPS}

\end{document}